\documentclass[a4paper,prx,aps,twocolumn,preprintnumbers,amsmath,amssymb,superscriptaddress]{revtex4-2}

\usepackage[utf8]{inputenc}
\usepackage[T1]{fontenc}
\usepackage{lmodern}
\usepackage{graphicx}
\usepackage{dcolumn}
\usepackage{bm}
\usepackage{textcomp}
\usepackage{float}

\usepackage[normalem]{ulem}
\usepackage{ifpdf}
\usepackage[squaren,Gray]{SIunits}
\usepackage{amsmath}

\ifpdf
\usepackage{epstopdf}
\usepackage[pdftex,unicode,pdfstartview={FitH},pdfborder={0 0 0}]{hyperref}
\usepackage{hypcap}
\else
\usepackage[hypertex]{hyperref}
\fi
\hypersetup{
    bookmarksnumbered = true,
    colorlinks = true, linkcolor = darkblue,
    citecolor = darkblue, filecolor = darkblue,
    menucolor = darkblue, urlcolor = black
}

\usepackage[utf8]{inputenc}
\usepackage{graphicx}% Include figure files
\usepackage{dcolumn}% Align table columns on decimal point
\usepackage{bm}% bold math

\usepackage{times}
\usepackage{amsmath}
\usepackage{amsfonts}
\usepackage{amssymb}
\usepackage{textcomp}
\usepackage{color}
\usepackage[normalem]{ulem}
\definecolor{red}{rgb}{1,0,0}
\definecolor{blue}{rgb}{0,0,1}
\definecolor{darkred}{rgb}{0.6,0,0}
\definecolor{darkblue}{rgb}{0,0,.6}
\definecolor{darkgreen}{rgb}{0,0.5,0}
\definecolor{grey}{rgb}{0.5,0.5,0.5}

\usepackage{titlesec}           % Change title format
\titleformat{\subsection}
{\bfseries} % format
{}          % label
{0.0cm}     % separation between label and body
{}          % code preceding title body
[]          % code following title body

\begin{document}

%\preprint{APS/123-QED}

\title{Imaging lattice reconstruction in homobilayers and heterobilayers\\ of transition metal dichalcogenides}

\author{Anna Rupp}
\def\LMU{Fakult\"at f\"ur Physik, Munich Quantum Center (MQC),
  and Center for NanoScience (CeNS),
  Ludwig-Maximilians-Universit\"at M\"unchen,
  Geschwister-Scholl-Platz 1, 80539 M\"unchen, Germany}
\affiliation{\LMU}

\author{Jonas G\"oser}
\affiliation{\LMU}

\author{Zhijie Li}
\affiliation{\LMU}

\author{Ismail Bilgin}
\affiliation{\LMU}

%\author{Kenji Watanabe}
%\def\RCFM{Research Center for Functional Materials, National Institute for Materials Science, 1-1 Namiki, Tsukuba 305-0044, Japan}
%\affiliation{\RCFM}

%\author{Takashi Taniguchi}
%\def\Japan{International Center for Materials Nanoarchitectonics,\\
%National Institute for Materials Science, 1-1 Namiki, Tsukuba 305-0044, Japan}
%\affiliation{\Japan}

\author{Anvar Baimuratov}
\affiliation{\LMU}

\author{Alexander H\"ogele}
\def\MCQST{Munich Center for Quantum Science and Technology (MCQST), Schellingtr. 4, 80799 M\"unchen, Germany}
\affiliation{\LMU}
\affiliation{\MCQST}

\date{\today}

\begin{abstract}
Moir\'{e} interference effects have profound impact on the optoelectronic properties of vertical van der Waals structures. Here we establish secondary electron imaging in a scanning electron microscope as a powerful technique for visualizing registry-specific domains in vertical bilayers of transition metal dichalcogenides with common moir\'e phenomena. With optimal parameters for contrast-maximizing imaging of high-symmetry registries, we identify distinct crystal realizations of WSe$_2$ homobilayers and MoSe$_2$-WSe$_2$ heterobilayers synthesized by chemical vapor deposition, and demonstrate ubiquitous lattice reconstruction in stacking-assembled bilayers with near parallel and antiparallel alignment. Our results have immediate implications for the optical properties of registry-specific excitons in layered stacks of transition metal dichalcogenides, and demonstrate the general potential of secondary electron imaging for van der Waals twistronics.
\end{abstract}

\maketitle
%\clearpage

%\section{Introduction}
%\vspace{-11pt}
Interference effects in vertical stacks of twisted two-dimensional crystal lattices induce moir\'{e} patterns, with rich consequences for charge carrier transport in the emergent mini-bands of the electronic band structure. As such, moir\'e effects give rise to peculiar transport phenomena in twisted homobilayer stacks of transition metal dichalcogenides (TMDs) \cite{wu2019topological,Wang2020a,Ghiotto2021} and also strongly affect the optoelectronic properties of TMD heterobilayers \cite{regan2020mott,tang2020simulation,Xu2020,Huang2021} with phenomena ranging from moir\'{e} intralayer \cite{yu2017moire, wu2017topological,zhang2018moire,zhao2023excitons} and 
interlayer \cite{yu2017moire,WuExciton2018,Tran2019,Jin2019,Seyler2019,forg2021moire,zhao2023excitons} excitons or hybrid excitons \cite{alexeev2019resonantly,hsu2019tailoring,shimazaki2020strongly,brem2020hybridized}. The actual manifestation of transport and optical phenomena depends in leading order on the twist angle and lattice mismatch, but additional effects of lattice reconstruction can be quite substantial \cite{carr2018relaxation,Rosenberger2020,enaldiev2020stacking,Weston2020,Sung2020,McGilly2020, andersen2021excitons,Halbertal2021,Shabani2021,Enaldiev2021,Weston2022,zhao2023excitons}. Ultimately, the details of moir\'e lattices and their reconstruction are essential for the interpretation of observed phenomena.  

Established imaging techniques with required resolution include conductive atomic force microscopy \cite{Rosenberger2020, Weston2020}, scanning tunneling microscopy \cite{Shabani2021,Halbertal2021,Li2023}, scanning transmission electron microscopy \cite{Weston2020,Li2023} and dark-field transmission electron microscopy \cite{alden2013strain,Rosenberger2020}. Providing high spatial resolution down to the limit of single atoms often involves extensive sample preparation methods. In contrast, scanning electron microscopy (SEM) combines comparably high spatial resolution with simple sample preparation. Moreover, SEM can provide complementary information such as elemental layer composition \cite{rupp2022energy} or crystallographic orientation \cite{ashida2015crystallographic} in secondary electron imaging, revealing reconstruction patterns in WSe$_2$ homobilayers with near-parallel alignment \cite{andersen2021excitons}.
 
In the following, we demonstrate the adoption and optimization of secondary electron imaging in a standard SEM for the visualization of domains with distinct atomic registries emerging in vertical TMD bilayers due to lattice reconstruction. To this end, we fabricate on the one hand TMD homobilayers and heterobilayers by stamping-assemblies of monolayers obtained from chemical vapor deposition (CVD), and on the other hand CVD-grown vertical homo- and heterobilayers, each with parallel and antiparallel alignment. With optimized operation conditions and numerical simulations, we identify domains of different atomic registries by their contrast in secondary electron imaging. We observe characteristic features of mesoscopic reconstruction, and find two distinct realizations of stable registries for parallel bilayer stacks grown directly by CVD-synthesis.

%%%%%%%%%%%%%%%%%%%%% Fig 1 %%%%%%%%%%%%%%%%%%%%%%%%%%%%%%%%%%%%%%%%
\begin{figure*}[t]
\centering
\includegraphics[scale=0.96]{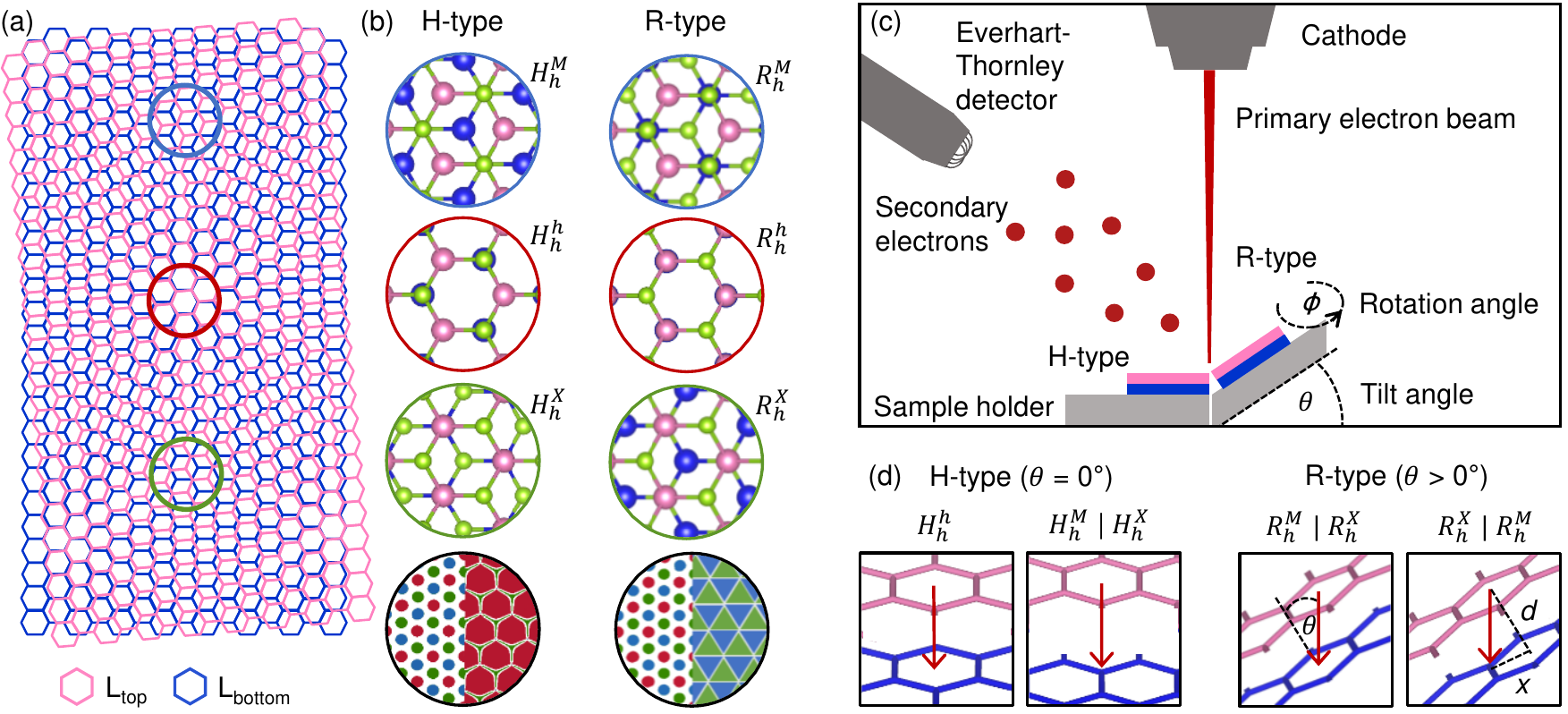}
\caption{(a) Schematics of moir\'{e} interference pattern in twisted bilayers, with points of three high-symmetry stackings indicated by colored circles. (b) Left (right) panel: Top view of the corresponding high-symmetry stackings $H^M_h$, $H^h_h$ and $H^X_h$ ($R^M_h$, $R^h_h$ and $R^X_h$) in antiparallel H-type bilayer (parallel R-type bilayer). The schematics in the bottom panel illustrate the rigid-lattice moir\'{e} pattern in the left halves of the circles which transforms upon reconstruction into hexagonal $H^h_h$ domains (triangular $R^M_h$ and $R^X_h$ domains) shown in the right halves. (c) Principle of reconstruction pattern imaging with a scanning electron microscope (not to scale): interactions between primary electrons and bilayer stack give rise to secondary electron emission detected with an Everhart-Thornley detector. The detection yield depends on the packing density of the incoming beam with respect to the bilayer. H-type imaging is performed at normal incidence, whereas R-type imaging requires a sample tilt $\theta$ and rotation angle $\phi$ for optimal contrast. (d) Side view of H-type imaging: At normal incidence channelling occurs only in the $H^h_h$ stacking. Side view of R-type imaging: Tilting the sample to $\theta$ results in channelling for only one stacking.}
\label{stackings}
\end{figure*}
%%%%%%%%%%%%%%%%%%%%%%%%%%%%%%%%%%%%%%%%%%%%%%%%%%%%%%%%%%%%%%%%%%%%%%%%%%%%%%%%%%%%%%%%

In ideal moir\'e superlattices formed by two hexagonal lattices upon vertical stacking with finite twist or lattice mismatch as indicated schematically in Fig.~\ref{stackings}(a), the characteristic moir\'{e} superlattice constant is given by  $\textit{L}_{M}(\delta)=a_1/{\sqrt{1+\left(a_1/a_2\right)^2 - 2(a_1/a_2)\cos \delta}}$, with the lattice constants of the two layers $a_1$ and  $a_2$, and the relative twist angle $\delta$ (modulo 60$\degree$) \cite{Moire_wavelength_Hermann}. Within one moir\'{e} supercell, three high-symmetry atomic registries stand out, illustrated by colored circles in Fig.~\ref{stackings}(b) - for heterobilayer stacks of non-centrosymmetric TMD monolayers: $H^M_h$, $H^h_h$ as well as $H^X_h$, and $R^M_h$, $R^h_h$ as well as $R^X_h$ stackings for near antiparallel (H-type) and parallel (R-type) alignment, respectively. The superscript and subscript refer to the electron and hole layer respectively, with M, X, and h denoting the transition metal atom, the chalcogen atom, and the hexagon center \cite{yu2017moire, baek2020highly}.

In the rigid-lattice moir\'{e} limit, all three stackings recur upon lateral translation as in the left halves of the bottom circles in Fig.~\ref{stackings}(b) for both H- and R-type stacks. In non-rigid bilayers which allow for finite atom displacement, however, the areas of energetically favorable $H^h_h$, $R^M_h$ and $R^X_h$ atomic registries are maximized at the expense of other stackings to give rise to periodically reconstructed patterns \cite{carr2018relaxation} that differ for H- and R-type stacks shown in the right halves of the bottom circles in Fig.~\ref{stackings}(b), respectively. In theory, periodic reconstruction yields hexagonal $H^h_h$ domains in H-type and alternating triangular $R^M_h$ and $R^X_h$ domains in R-type stacks \cite{carr2018relaxation,enaldiev2020stacking,Enaldiev2021}, with domain sizes proportional to the moir\'{e} period length $L_M$ ranging from a few to a few tens of nanometers. In practice, however, regions of regular periodic reconstruction are rather limited, and irregular reconstructed landscapes with variations on mesoscopic length scales prevail instead \cite{Rosenberger2020,Weston2020,Sung2020,McGilly2020, andersen2021excitons,Halbertal2021,Shabani2021,Weston2022,zhao2023excitons}. 

%\section{Experimental methods and results} 
%\vspace{-11pt}
Secondary electron detection in SEM represents a powerful experimental method to visualize reconstructed crystal morphology on length scales down to the spatial resolution limit of low-energy primary electrons. It is based on the detection of secondary electrons generated by inelastic scattering of the primary electron beam with core or valance electrons of the sample, with the principle of operation shown in the schematics of Fig.~\ref{stackings}(c). Given kinetic energies of secondary electrons below 50\:eV, their escape depth is confined to near-surface regions. A conventional Everhart-Thornley detector inside the SEM records the secondary electron yield at each scan position and converts it into a grey value of an image pixel. The more interactions between primary electrons and surface-layer atoms occur, the more secondary electrons are generated and counted, corresponding to a brighter image pixel \cite{reimer2000scanning,goldstein2017scanning}. Due to distinct stackings, different experimental conditions are required for optimal imaging of H- and R-type bilayers, as indicated in Fig.~\ref{stackings}(c) and (d).

%%%%%%%%%%%%%%%%%%%%%%%%%%%%%%%%%%%%%%%% Fig 2 %%%%%%%%%%%%%%%%%%%%%%
\begin{figure*}[t]
\centering
\includegraphics[scale=0.96]{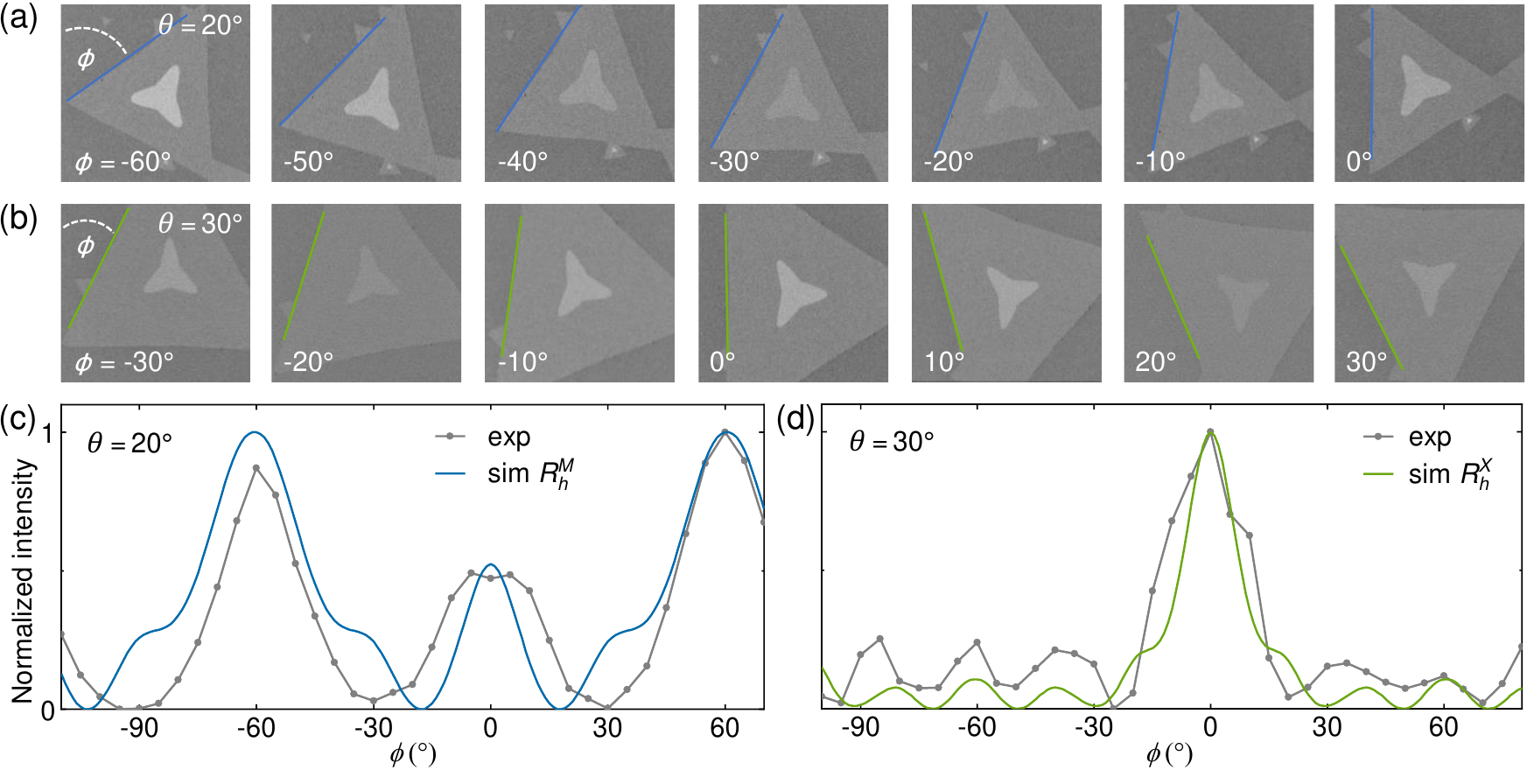}
\caption{(a) and (b),  Secondary electron images of two CVD-grown R-type WSe$_2$ homobilayers tilted to $\theta = 20\degree$ and $30\degree$, respectively, for different values of the rotation angle $\phi$ in steps of $10\degree$. (c) and (d), Corresponding contrast modulation in experiment (grey data in steps of $5\degree$) and simulations (blue and green solid lines for $R^M_h$ or $R^X_h$ stackings, respectively) as a function of the rotation angle $\phi$. Note that the two homobilayers synthesized in the same CVD process differ in their R-type stacking.}
\label{theta}
\end{figure*}
%%%%%%%%%%%%%%%%%%%%%%%%%%%%%%%%%%%%%%%%%%%%%%%%%%%%%%%%%%%%%%%%%%%%%

To begin with, we consider the concept of optimal channelling conditions for secondary electrons in H- and R-type stacks. For H-type stacks, optimal channelling is realized at normal incidence of the incoming beam (corresponding to a tilt angle $\theta = 0\degree$), as illustrated in Fig.~\ref{stackings}(d). This configuration gives rise to maximum contrast between domains of $H^h_h$ atomic registry and $H^M_h$ or $H^X_h$ domains with higher packing density from the perspective of the incoming beam. The difference in the packing density for normal incidence in turn results in different secondary electron yields and hence different contrast in the SEM image. The intuitive model of classical electrons suggests less interactions between primary electrons and lattice atoms for the channelling condition in the left panel of Fig.~\ref{stackings}(d), where the electron trajectory is indicated to pass the hexagon centers of both layers in the $H^h_h$ case as opposed to encountering the hexagon corner of the bottom layer in $H^M_h$ or $H^X_h$ stacking. The wave nature of low-energy electrons, however, leads to diffraction, which in the channelling case gives rise to more backscattering of primary electrons and thus an increased generation of secondary electrons upon reversed propagation of backscattered electron to the surface \cite{joy1982electron}. Overall, the secondary electron yield is thus higher for the channelling condition \cite{ashida2015crystallographic}, and $H^h_h$ domains appear bright in SEM imaging while $H^M_h$ and $H^X_h$ stacks are dark. %\cite{ashida2015crystallographic} is for SiC channelling conditions.

Maximum contrast at normal incidence is not expected for bilayers with R-type parallel alignment and reconstruction into $R^M_h$ and $R^X_h$ domains with similar packing densities. In contrast to optimal conditions at normal incidence for H-stacks, a finite tilt of the sample at angle $\theta$ is required to obtain channelling with mutually contrasting generation yields of secondary electrons by domains of $R^M_h$ and $R^X_h$ atomic registries, as illustrated in the right panel of Fig.~\ref{stackings}(d). Again, channelling results in a higher secondary electron yield and thus a brighter image pixel. The optimal tilt angle can be calculated from the channelling condition shown in Fig.~\ref{stackings}(d) as $\theta = \arctan{(x/d)}$ with the lateral distance of the beam passage through the bilayers $x$ and the vertical interlayer distance $d$. Moreover, for samples with finite tilt angle $\theta$, the secondary electron yield, and hence the image grey value, is also a function of the rotation angle $\phi$. This in turn allows to discriminate between $R^M_h$ and $R^X_h$ stacked domains.

%%%%%%%%%%%%%%%%%%%%%%%%%%%%%%%%% Fig 3 %%%%%%%%%%%%%%%%%%%%%%%%%%%%
\begin{figure*}[t]
\centering
\includegraphics[scale=0.96]{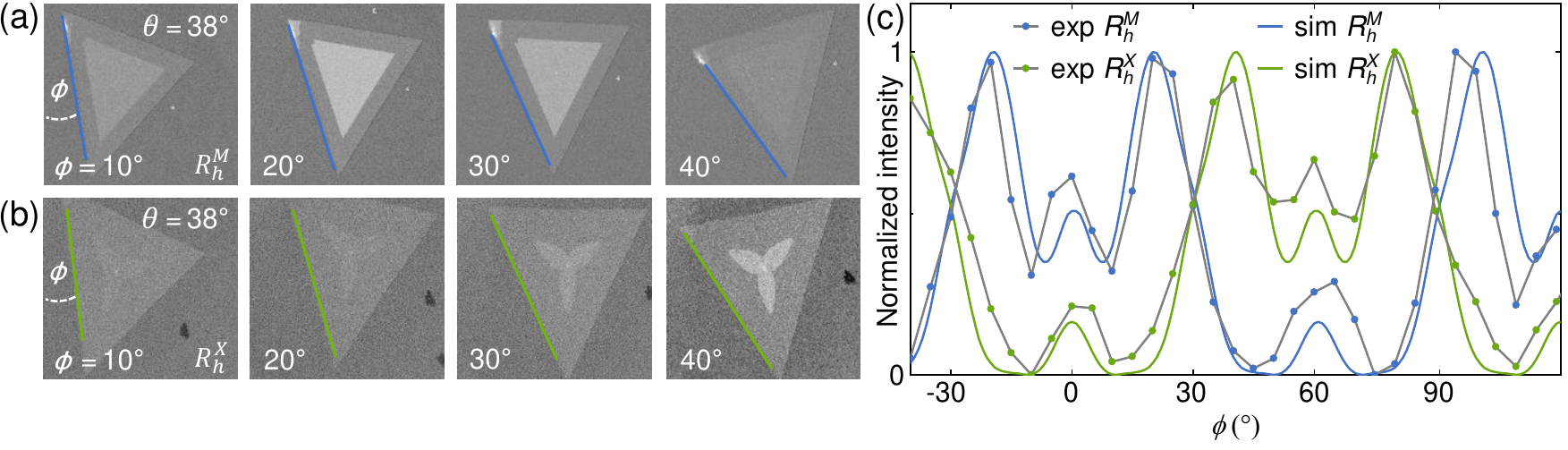}
\caption{(a) and (b), SEM images of two CVD-grown R-type WSe$_2$ bilayers both tilted to $\theta$ = 38$\degree$ and inspected at the same $\phi$-angles. The reversed contrast indicates opposite stacking orders. The corresponding contrast modulations for both bilayers as a function of the rotation angle are shown by the grey lines in (c). Simulations of $R^M_h$ and $R^X_h$ WSe$_2$ bilayers for $\theta$ = 38$\degree$ tilt as a function of the rotation angle assign the $R^M_h$ and $R^X_h$ stacking to the top and bottom bilayer in (a) and (b), respectively. Owing to the 60$\degree$ shifted modulation between $R^M_h$ and $R^X_h$ stackings only four angles exhibit maximum contrast between the two stackings.}
\label{phi}
\end{figure*}
%%%%%%%%%%%%%%%%%%%%%%%%%%%%%%%%%%%%%%%%%%%%%%%%%%%%%%%%%%%%%%%%%%%%%%

The dependence of the imaging contrast on the tilt and rotation angles, $\theta$ and $\phi$, are shown in Fig.~\ref{theta} for CVD-grown WSe$_2$ homobilayers. The use of CVD-bilayers is particularly advantageous, as the crystallographic orientation is directly evident from the relative ordination of the inner and outer triangular-shaped crystal layers. In Fig.~\ref{theta}(a) and (b), the two R-type bilayers, synthesized in the same growth process, feature parallel alignment of top (bright) and bottom (dark) triangles (with deviations from triangular shapes in the top layer). Both bilayer crystals exhibit contrast modulation in Fig.~\ref{theta}(a) and (b) as a function of rotation angle $\phi$ for tilt angles $\theta = 20\degree$ and $30\degree$, respectively. The contrast modulation is cyclic in $\phi$ (with $\phi$ = 0$\degree$ defined with respect to the tilt axis chosen collinear with the triangle edge that is closest to the secondary electron detector, and positive values corresponding to counterclockwise rotation), as evident from the normalized data in Fig.~\ref{theta}(c) and (d). At each point, the grey value of the substrate was subtracted from the grey value of the bilayer, and the contrast was normalized to the cyclic maximum. 

Even though both bilayers of Fig.~\ref{theta}(a) and (b) are clearly R-type, they exhibit different contrast evolution in secondary electron imaging as a function of $\theta$ and $\phi$. To identify the respective registry configurations, we performed Monte-Carlo simulations of the secondary electron yield as a function of tilt and rotation angle for all high-symmetry stackings of R- and H-type bilayers \cite{andersen2021excitons} (see Methods for details). For R-type, the simulations shown by solid lines in Fig.~\ref{theta}(c) and (d) reproduce the experimental contrast modulations with very good agreement and lead to the following main conclusions. Most generally, higher tilt angles yield larger contrast modulations due to increasing interaction volume with tilt. Moreover, the contrast modulation under $\phi$-rotation, with a periodicity of $120\degree$ due to crystal symmetry, is characteristic for a given tilt angle $\theta$ and registry, discriminating between $R^M_h$ and $R^X_h$ atomic registries of the homobilayers in Fig.~\ref{theta}. To the best of our knowledge, this observation is the first to identify CVD-grown homobilayers in the contrasting limits of fully reconstructed $R^M_h$ and $R^X_h$ crystal stackings.

%%%%%%%%%%%%%%%%%%%%%%%% Fig 4 %%%%%%%%%%%%%%%%%%%%%%%%%%%%%%%%%%%%%
\begin{figure*}[t]
\centering
\includegraphics[scale=0.96]{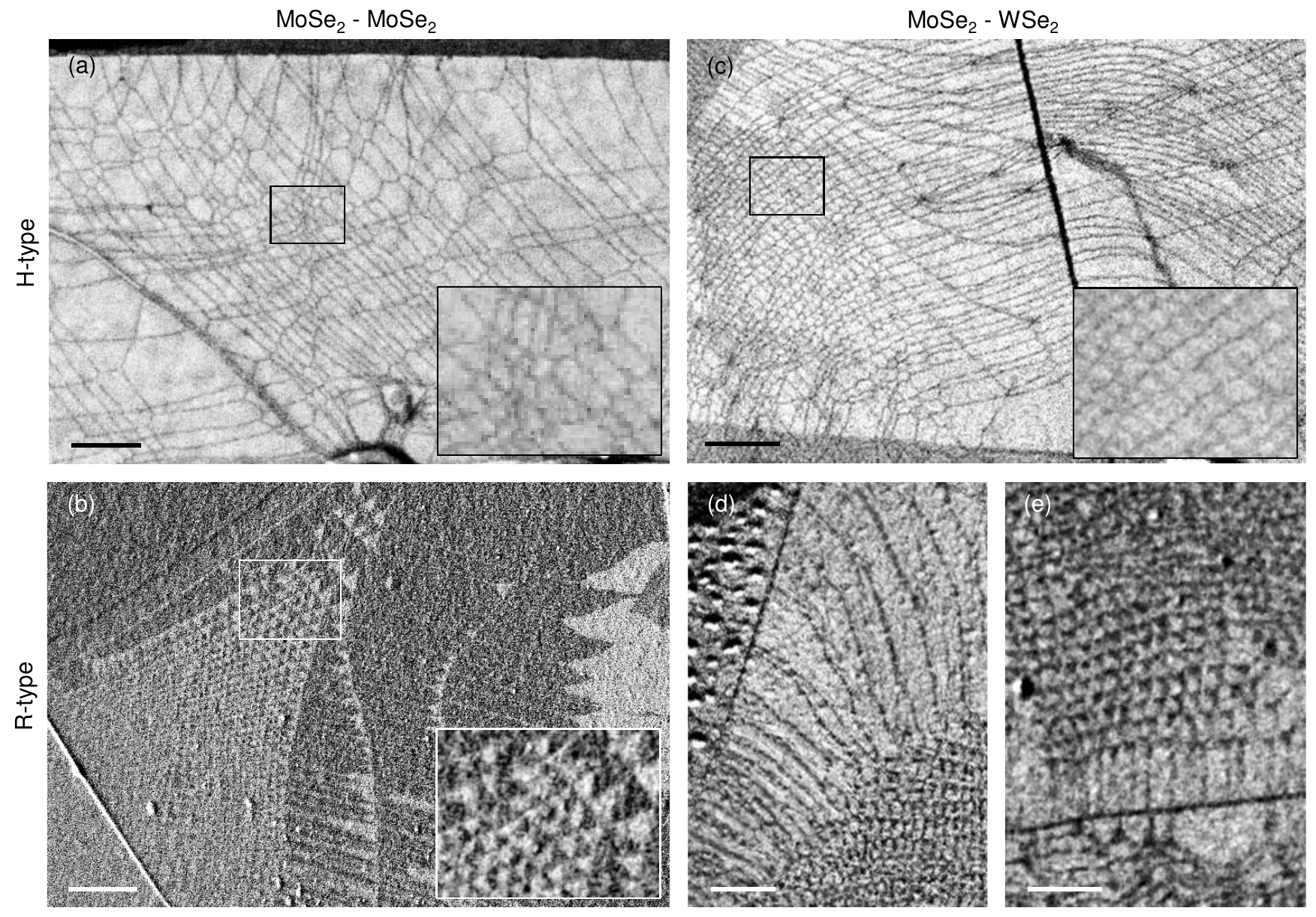}
\caption{(a) and (b), Secondary electron images of twisted H-type ($\delta=179.6\degree$) and R-type ($\delta=0.2\degree$) MoSe$_2$ homobilayers fabricated as stamping-assembly of CVD-grown monolayers. Bright domains in (a) correspond to $H^h_h$, whereas bright and dark domains in (b) correspond to $R^M_h$ and $R^X_h$ stackings. The insets show zooms to regions with nanoscopically reconstructed domains in hexagonal and triangular tilting. (c) Same but for a twisted H-type ($\delta=179.7\degree$) MoSe$_2$-WSe$_2$ heterobilayer. (d) and (e) Two regions of a twisted R-type ($\delta=0.3\degree$) MoSe$_2$-WSe$_2$ heterobilayer. All scale bars are $500$~nm. The images were recorded at $\theta=0\degree$ and $38\degree$ for H- and R-type stacks, respectively, with a rotation angle $\phi=-20\degree$ in (b), (d) and (e).}
\label{images}
\end{figure*}
%%%%%%%%%%%%%%%%%%%%%%%%%%%%%%%%%%%%%%%%%%%%%%%%%%%%%%%%%%%%

%%%%%%%%%%%%%%%%%%%% Fig 5 %%%%%%%%%%%%%%%%%%%%%%%%%%%%%%%%%%%%%
\begin{figure*}[t]
\centering
\includegraphics[scale=0.96]{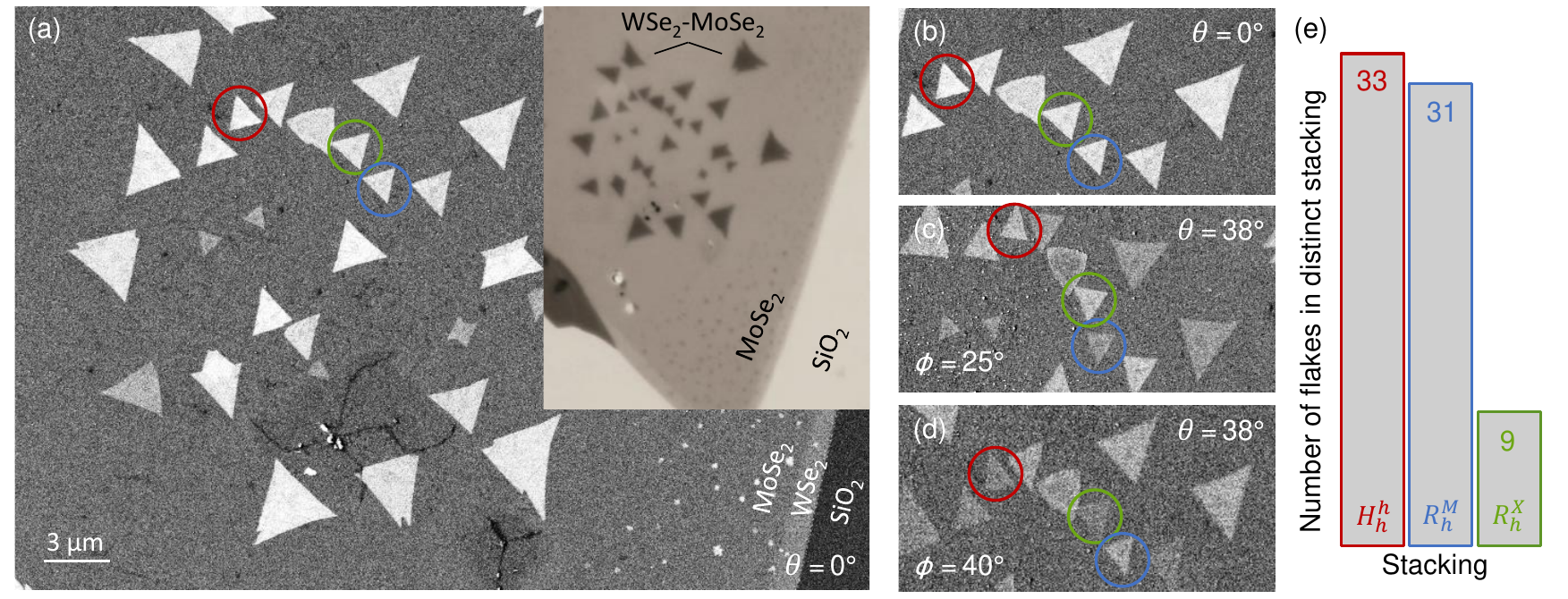}
\caption{(a) Secondary electron image at normal incidence ($\theta=0\degree$) and optical micrograph (top right inset) of CVD-grown WSe$_2$-MoSe$_2$ heterostructures formed by a large bottom MoSe$_2$ monolayer and several WSe$_2$ triangular monolayers on top. Three representative triangles with different atomic registries are encircled in red, green and blue. (b), (c) and (d) Zooms to the region with the three triangles at normal incidence, and at $\phi=25\degree$ and $40\degree$ rotation for a tilt of $\theta=38\degree$. (e) Histogram count of different heterobilayer flakes with distinct atomic registries from several CVD-grown heterostacks. Note that due to reversed layer order of the CVD-grown heterostack the relative contrast of R-type registries is reversed.}
\label{correlation}
\end{figure*}
%%%%%%%%%%%%%%%%%%%%%%%%%%%%%%%%%%%%%%%%%%%%%%%%%%%%%%%%%%%%%%%%%%%%%%%%%%%%%%%%

This contrasting behavior in secondary electron imaging is highlighted in Fig.~\ref{phi} on another pair of WSe$_2$ homobilayer flakes obtained in the same synthesis run. For the same tilt angle of $\theta=38\degree$, which represents a balanced optimum between increased modulation contrast and reduced spatial resolution upon tilt, we clearly observe the limiting cases of $R^M_h$ and $R^X_h$ reconstructed domains for the homobilayer crystals in Fig.~\ref{phi}(a) and (b) for the same rotation angles $\phi$ = 10$\degree$, $20\degree$, $30\degree$ and $40\degree$. Consistently, the respective normalized contrasts in Fig.~\ref{phi}(c) differ substantially for the two registries upon $\phi$-rotation, maximizing the registry-discriminating contrast for specific $\phi$ values (20$\degree$, 40$\degree$, 80$\degree$ and 100$\degree$). 

Equipped with this understanding and optimized operation parameters, we utilized secondary electron imaging to visualize reconstruction effects in stacking-assembled MoSe$_2$ homobilayers and MoSe$_2$-WSe$_2$ heterobilayers, each near antiparallel and parallel alignment. The samples were fabricated with a stamping-based direct pick up method \cite{purdie2018cleaning} by placing a CVD-grown monolayer onto another monolayer CVD-grown on a Si/SiO$_2$ substrate. The relative orientation close to $0\degree$ or $180\degree$ alignment was again facilitated by the triangular shapes of monolayer crystals, with an accuracy of $\pm 0.3\degree$ as determined a posteriori with SEM. Using optimized parameters discussed above, secondary electron imaging was performed at $\theta=0\degree$ and $38\degree$ for H- and R-type stacks, respectively.  

Overall, the morphologies of mechanically stacked homobilayer MoSe$_2$ in Fig.~\ref{images}(a) and (b) and heterobilayer MoSe$_2$-WSe$_2$ in Fig.~\ref{images}(c), (d) and (e) are consistent with lattice reconstruction on mesoscopic length scales  \cite{Rosenberger2020,Weston2020,Sung2020,McGilly2020, andersen2021excitons,Halbertal2021,Shabani2021,Weston2022,zhao2023excitons}. In secondary electron images of H-type stacks, recorded at zero tilt and shown in Fig.~\ref{images}(a) and (c), bright regions correspond to $H^H_h$ stacking. The images of R-type stacks in Fig.~\ref{images}(b), (d) and (e), recorded at $\theta=38\degree$ and $\phi=-20\degree$, exhibit maximum contrast between bright $R^M_h$ and dark $R^X_h$ stackings. All images show local variations in the domain pattern, with relatively large domains frequently found near edges or folds, and rather periodic patterns with domain sizes well below $100$~nm observed predominantly in the sample core. In accord with theoretical anticipation of lattice reconstruction from moir\'e to periodic domains with hexagonal and triangular tiling in H- and R-type stacks \cite{carr2018relaxation,enaldiev2020stacking}, we frequently observed domains with the corresponding geometries, as emphasized in the insets of Fig.~\ref{images}(a), (b) and (c), as well as in Fig.~\ref{images}(d). 

Remarkably, the inspection of R-type homobilayers as in Fig.~\ref{images}(b) indicates that both bright and dark domains of $R^M_h$ and $R^X_h$ stackings emerge upon large-scale reconstruction \cite{carr2018relaxation, enaldiev2020stacking}, whereas heterobilayers tend to reconstruct preferentially into bright $R^M_h$ domains at the expense of dark $R^X_h$ domains as in Fig.~\ref{images}(d) and (e). This suggests an energetic imbalance in the competition between $R^M_h$ and $R^X_h$ registries in heterobilayers that is absent in homobilayers, in accord with DFT calculations for MoSe$_2$-WSe$_2$ heterobilayer with a slightly favored $R^M_h$ stacking \cite{Rosenberger2020,Li2023}.

The competition between domains of $R^M_h$ and $R^X_h$ registries in large-scale reconstruction is even more evident in CVD-grown MoSe$_2$-WSe$2$ heterobilayers. In Fig.~\ref{correlation}(a), we show optical and secondary electron images of a MoSe$_2$-WSe$2$ heterostack formed by small WSe$_2$ triangular monolayers on top of a large MoSe$_2$ monolayer triangle. The zoom to the secondary electron image at normal incidence ($\theta=0\degree$) in Fig.~\ref{correlation}(b) singles out a section with three representative triangles encircled in red, green and blue, which we inspected at a tilt of $\theta=38\degree$ for two rotation angles $\phi$ of $25\degree$ and $40\degree$ in Fig.~\ref{correlation}(c) and (d), respectively. 

First, we point out similar grey-scale contrasts observed at $\theta=0\degree$ for all three triangles in Fig.~\ref{correlation}(a) and (b). However, at finite tilt of $\theta=38\degree$, they interchange their relative brightness upon rotation from $\phi=25\degree$ to $40\degree$ in Fig.~\ref{correlation}(c) and (d), respectively. This behavior, combined with our understanding of registry-specific contrasts from simulations, identify the triangles marked in red, green and blue as exhibiting complete lattice reconstruction into $H^h_h$, $R^X_h$, and $R^M_h$ domains. The mutual exclusion of the latter two, in particular, is a hallmark of the energy-driven competition between two near-optimal stackings in R-type MoSe$_2$-WSe$_2$ heterostacks for large-area domains. 

The observation of nearly complete reconstruction of CVD-grown WSe$_2$-MoSe$_2$ heterobilayers (except for the inner cores as discussed previously elsewhere \cite{Li2023}) into one exclusive registry is remarkable, and was observed as a robust feature on more than $70$ heteroflakes. The distribution of distinct stackings for H- and R-type heterostacks is shown in Fig.~\ref{correlation}(e), featuring $H^h_h$ as the singular stacking for H-type stacks, and the $R^M_h$ stacking outcompeting the $R^X_h$ stacking with a ration of $\sim 3$. This distribution provides further support into the energetic preference of MoSe$_2$-WSe$_2$ with regard to reconstruction into $R^M_h$ domains as predicted by DFT calculations \cite{Rosenberger2020,Li2023}.

%\section{Conclusions}
%\vspace{-11pt}
In summary, we reported secondary electron imaging of reconstructed homo- and heterobilayers, each with parallel and antiparallel alignment. Whereas registry contrast in antiparallel bilayers is readily achieved with low beam energy and high resolution at normal incidence, discrimination of different domains in parallel stacks requires optimization of both tilt and rotational angles. With optimized parameters and numerical simulations, we identified lattice reconstruction on mesoscopic length scales as the predominant effect in the formation of diverse morphologies in stamping-assembled MoSe$_2$ homobilayers and MoSe$_2$-WSe$_2$ heterobilayers near parallel and antiparallel alignment. For CVD-grown WSe$_2$ homobilayers and WSe$_2$-MoSe$_2$ heterostructures, our imaging technique revealed complete reconstruction into domains of a single registry, with mutually exclusive $R^M_h$ and $R^X_h$ stacking configurations in R-type flakes and the tendency of $R^M_h$ domain predominance. Our findings have immediate consequences for the optical properties of homobilayer \cite{Li2022} and heterobilayer stacks \cite{zhao2023excitons,Li2023} of semiconducting TMDs, and can be generalized to the entire class of van der Waals heterostructures with small lattice mismatch and twist angles.\\

%\newpage
%\vspace{6pt}
%\section{Methods}
\noindent \textbf{METHODS}\\

\noindent \textbf{Sample fabrication:} WSe$_2$ homobilayers as in Fig.~\ref{theta} and Fig.~\ref{phi} were synthesized by a one-step CVD growth as detailed in Ref.~\cite{Li2022}. For the MoSe$_2$-MoSe$_2$ and MoSe$_2$-WSe$_2$ bilayers in Fig.~\ref{images}, triangular single-crystal monolayers of MoSe$_2$ and WSe$_2$ were obtained separately from CVD synthesis. The MoSe$_2$ layers were picked up with an adhesive polycarbonate stamp at a temperature of $145\degree$C. The monolayer alignment to $0\degree$ (R-type) or $180\degree$ (H-type) was guided by straight triangle edges, with alignment precision limited to below $0.3\degree$. The assembled stacks were successively released from the stamp onto a SiO$_2$/Si target substrate at a temperature of $180\degree$C, then soaked in chloroform solution for $20$~min to remove polycarbonate residues, cleaned by acetone and isopropanol and annealed at $200\degree$C under ultrahigh vacuum for $12$~hours. The WSe$_2$-MoSe$_2$ heterobilayers in Fig.~\ref{correlation} were grown by a two-step CVD process detailed in Ref.~\cite{Li2023}.

\vspace{8pt}
\noindent \textbf{Secondary electron imaging:} Secondary electron imaging was performed in a Raith-eLine SEM with an electron beam energy of $1.4$ and $1.0$~keV for H- and R-type stacks, respectively, at a working distance of $4.5$~mm and an aperture of $30$~\textmu m. Carbon deposition during SEM-imaging \cite{goldstein2017scanning,reimer2000scanning} was observed to reduce progressively the visibility of the registry-specific contrast. We note that the stacking nomenclature is relevant for the respective contrast in secondary electron imaging of heterobilayer registries. For the MoSe$_2$-WSe$_2$ heterostructures studies in this work, the samples in Fig.~\ref{phi}(b) correspond to vertical stacks of bottom WSe$_2$ and top MoSe$_2$ monolayers, as shown schematically in Fig.~\ref{stackings}(a) and (b) and used to introduce the registry nomenclature. The CVD-grown heterostack of Fig.~\ref{correlation}, on the other hand, features the reversed layer order and thus opposite contrast behavior for a registry nomenclature with the subscript and superscript referring to the bottom and top layer, respectively. For stacks with bottom MoSe$_2$ and top WSe$_2$ monolayers, this implies that bright contrast at $\phi=40\degree$ corresponds to the R$^M_h$ stacking.

%\vspace{6pt}
\noindent \textbf{Numerical simulations:} Monte-Carlo simulations of the secondary electron yield as a function of tilt and rotation angle \cite{andersen2021excitons} were performed for all six registry configurations of a WSe$_2$ bilayer with lattice constant $a_0=3.28$~\AA~ and interlayer distance $d=6.60$~\AA~\cite{he2014stacking} subjected to $49152$ parallel incident electron trajectories per unit cell. We accounted for electron-electron interactions between the incident and bound electrons by a cylindrical cross-section of $r=4a$ around each atom, with the nearest-neighbor interatomic distance $a=a_0/\sqrt{3}$. The scattering probability was calculated as $A (b/B)^{-2}$, where $A$ specifies the secondary electron yield proportional to the number of covalently bound electrons, $B$ is  proportional to the covalent radius, and the impact parameter $b$ is sensitive to the distance between the incoming electrons and the crystal atoms \cite{andersen2021excitons}. The constants $A$ and $B$ were determined from best-fits to experimental data with $A_{\mathrm{Se}} = 0.09$, $A_{\mathrm{W}} = 0.14$ and $B_{\mathrm{Se}} = 0.3$, $B_{\mathrm{W}} = 0.4$. The scattering probabilities were averaged over all trajectories at each $\phi$ for a given $\theta$. Finally, the scattering probability was subtracted from unity to obtain the experimentally observed inverse secondary electron yield as discussed above (i.e. high secondary electron yield for the channelling case).

%\section{Acknowledgements}
\noindent \textbf{ACKNOWLEDGEMENTS}\\
%\vspace{-8pt}

\noindent We thank P. Altpeter and C. Obermayer for assistance in the clean room. This research was funded by the European Research Council (ERC) under the Grant Agreement No.~772195 as well as the Deutsche Forschungsgemeinschaft (DFG, German Research Foundation) within the Priority Programme SPP~2244 2DMP and the Germany's Excellence Strategy EXC-2111-390814868 (MCQST). A.~R. acknowledges funding by the Munich Quantum Valley doctoral fellowship program within the Bavarian initiative "Hightech Agenda Bayern Plus". A.~S.~B. received funding from the European Union’s Framework Programme for Research and Innovation Horizon 2020 (2014–2020) under the Marie Skłodowska-Curie Grant Agreement No. 754388 (LMUResearchFellows) and from LMUexcellent, funded by the Federal Ministry of Education and Research (BMBF) and the Free State of Bavaria under the Excellence Strategy of the German Federal Government and the Länder. Z.~L. was supported by the China Scholarship Council (CSC), No. 201808140196. S.~Z. and I.~B. acknowledge support from the Alexander von Humboldt Foundation.

%\cleardoublepage
%\bibliographystyle{naturemag}
%\bibliography{Moire_bibliography_v3}

%

\end{document}